# On Reliability of Android Wearable Health Devices

Naixing Wang, Edgardo Barsallo Yi, Saurabh Bagchi
Purdue University
West Lafayette, IN, USA
{wang2489, ebarsall, sbagchi}@purdue.edu

## ABSTRACT
Wearable devices are increasingly being used for monitoring health signals and for fitness purposes with typical uses being calorie tracker, workout assistant, and step counter. Even though these wearables can measure many health signals (e.g. heart rate), they are still not perceived as highly accurate, relative to clinical monitoring devices. In this paper, we investigate the accuracy of heart monitor as included in two popular wearables – Motorola Moto 360 and the Apple Watch. We analyze the accuracy from a hardware and a software perspective and show the effects of body motion on the heart rate monitors based on the use of photoplethysmography (PPG) signals used in Android wearables. We then do a software reliability study of the Android Wear OS, on which many wearables are based, using fuzz testing.

## 1. INTRODUCTION
Most of the wearables devices rely on an optical sensor based on PPG to monitor the heart rate. However, PPG signals are highly contaminated by artifacts caused by movements of the subject. Such motion artifacts strongly interfere with accurate heart rate measurements, especially if the PPG signal is recorded from the wrist during intensive physical exercise. However, this is precisely the usage mode that is relevant to wearable smart devices, such as smartwatches.

Many related works have been proposed on motion artifacts analysis and noise reduction for heart rate monitoring [1, 2]. Specifically, [2] analyzes the accuracy of heart rate monitoring with PPG signal using different wavelengths of light (red, green, and blue). The results show that the green light PPG has a higher accuracy and a much larger signal to noise ratio and this has driven the choice made by vendors on most wearable devices.

In this paper, we analyze the reliability of the heart rate monitor of android wearables, such as the Motorola Moto 360 2nd Generation (Moto 360 G2). We compare its reliability against a perceived more reliable wearable and dominant product, Apple Watch 1st Generation, and use a clinical-grade medical fingertip pulse oximeter as base reference for the experiments. Contrary to smartwatches, which measure heart rate from the wrist, fingertip pulse oximeter tends to be more accurate due to a high capillary density in the tissue on fingertips.

Our work is the starting point to analyze the reliability of Android wearable devices in a data-driven manner, and follows in line with our prior work in analyzing the reliability of mobile operating systems [3]. The main contribution here are:
- A study showing error effects on the PPG signal due to body motion.
- A guideline to improve the accuracy of the heart rate monitor on Android Wear.

## 2. ACCURACY OF WEARABLES
The reliability study is focused on the heart rate monitor. We analyzed the accuracy of measures obtained from the heart rate sensor in Motorola Moto 360 G2, released in 2015 and considered today a leading-edge Android Wear smartwatch.

In the experiments, we compare the accuracy of the heart rate measurements between the Android and Apple smartwatches when the examinee is resting and during regular exercise; while a fingertip pulse oximeter is used as the golden reference. As described in the previous section, the main issue in measuring the Heart Rate (HR) using PPG signal is the motion artifacts, since they distort this signal. Typically, motion artifacts can be translated into accelerations in a 3D coordinate system.

Figure 1 shows the experimental results of the two different scenarios. In Figure 1(a), we compare the HR monitoring when the examinee is resting. The plot indicates that both devices report HR close to the golden reference when the examinee is resting. However, there is a non-negligible delay time for the readings of Moto 360 G2. Figure 1(b) shows the experimental results of the two wearable devices when the examinee is jogging at 5 miles/hr. In the case of the fingertip pulse oximeter, the HR measurements do not change during the exercise since the device does not to produce any measurement during body motion. Therefore, we only record the HR measurements before the exercise and after the exercise and the golden values are 67 BPM (beats per minute) and 93 BPM respectively. It can be observed that both smartwatches are not accurate due to the motion artifacts. Nevertheless, the measures from the Apple Watch are more accurate.

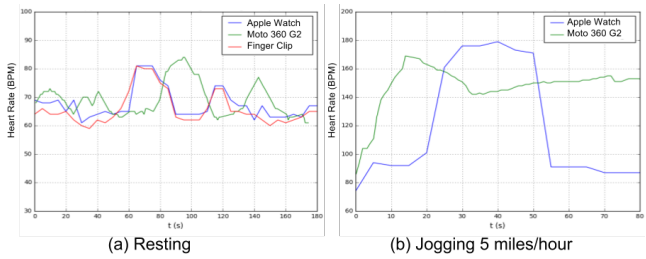

Figure 1. HR comparison between devices.

## 3. EFFECT OF MOTION ON HR SIGNAL

To investigate the relationship between the distorted PPG signal and the acceleration on each axis for the Moto 360, we follow similar procedures used in [1] to measure the PPG signal from the forehead of the subject. First, the examinee is asked to wave his arm/wrist left and right along the X-axis. The examinee is not allowed to move his arm/wrist along the other axes during the experiment. This is then repeated for the Y-axis and the Z-axis. To investigate the relationship between the PPG signal and the accelerometer signal during the regular exercise, in the fourth experiment, the examinee is asked to wave his arm/wrist as in regular walking.

Figure 2 shows the experimental results of the four scenarios. The experimental results indicate that the PPG signal is more correlated with x-axis acceleration when the disrupted PPG signal is delayed by 0.2s—the correlation coefficient is not very high though, it is 0.53 with a delay of 0.2s. This means the motion artifacts show up in the PPG measurement with a non-negligible time delay. In addition, the noise which distorts the PPG signal is mainly generated from the X-axis acceleration.

## 4. FUZZ TESTING OF ANDROID WEAR

To verify the reliability of Android Wear 1.0 operating system, we seek to apply the fuzz testing methodology, since this technique is able to identify most common errors and potential vulnerabilities quickly and cost effectively. For this purpose, Android Monkey UI/Application Exerciser tool was used to stress-test the built-in apps included in the Moto 360. Basically, the tool generates pseudo random stream of user events, which can be used to stress the app user interface. The experiment consisted of two stages. Initially, we ran 4,000 trials for each app. Then, for those apps which crashed in the first round, we ran an additional 10,000 trials. The experiments showed that the fitness apps (Google Fitness and Motorola Body) are the less reliable apps on the Android Wear. The Google Fitness app crashed once, while Moto Body app crashed 6 times. The crashes were due NullPointerException errors, basically for attempting to read a field with null value or invoke a method using a null reference object.

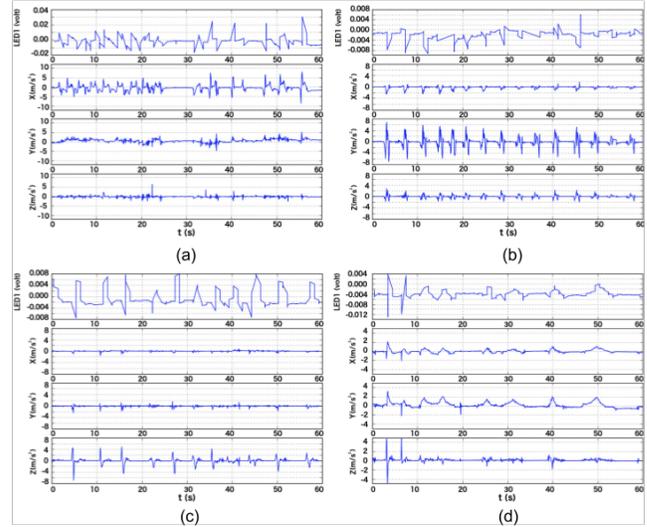

Figure 2. PPG signal against acceleration. In (a)-(c), the acceleration is only varied on each axis respectively. In (d), the acceleration is varied on all the axes.

## 5. CONCLUSIONS AND FUTURE WORK

In this paper, we presented an analysis of the accuracy of mobile heart rate monitoring using Android wearable device. The experimental results indicate that the heart rate monitoring is accurate during resting, but less accurate with body motion. This is because the PPG signal is affected by motion artifacts. In addition, the distorted PPG signal is more correlated to the x-axis acceleration. The motion artifacts also reflect a non-negligible delay time (~0.2s). Based on the experiments, the accuracy of mobile heart rate monitoring can be improved by discarding unreliable data and noise cancellation. We also showed initial results about the reliability of the health-related apps on Android Wear.

As a primary future work, we plan to extend our experiments to address the reliability on the Android Wear OS. We plan to build a fuzz testing framework capable of fuzzing the hardware sensors data and use it to check the reliability of the Google Fit API. The experiments will be conducted using a white box approach, instead of a black box, as was done for this work.